\documentclass[
superscriptaddress,
final,
reprint,
showkeys,
aps,
pre,
floatfix
]{revtex4-1}

\usepackage{nameref}
\usepackage[dvips]{graphicx}
\usepackage[utf8]{inputenc}
\usepackage{multirow,booktabs}
\usepackage{amsmath,amssymb,latexsym,MnSymbol,bm}
\usepackage{blkarray}

\usepackage[table]{xcolor}
\usepackage{xcolor}
\usepackage[%
  colorlinks=true,
  urlcolor=blue,
  linkcolor=blue,
  citecolor=blue
]{hyperref}
\usepackage{blindtext}
\usepackage{scalerel,stackengine}

\begin{document}

\title{Characterizing stochastic time series with ordinal networks}

\author{Arthur\ A.\ B.\ Pessa}
\email{arthur\_pessa@hotmail.com}
\affiliation{Departamento de F\'isica, Universidade Estadual de Maring\'a -- Maring\'a, PR 87020-900, Brazil}

\author{Haroldo V.\ Ribeiro}
\email{hvr@dfi.uem.br}
\affiliation{Departamento de F\'isica, Universidade Estadual de Maring\'a -- Maring\'a, PR 87020-900, Brazil}

\date{\today}

\begin{abstract}
Approaches for mapping time series to networks have become essential tools for dealing with the increasing challenges of characterizing data from complex systems. Among the different algorithms, the recently proposed ordinal networks stand out due to its simplicity and computational efficiency. However, applications of ordinal networks have been mainly focused on time series arising from nonlinear dynamical systems, while basic properties of ordinal networks related to simple stochastic processes remain poorly understood. Here, we investigate several properties of ordinal networks emerging from random time series, noisy periodic signals, fractional Brownian motion, and earthquake magnitude series. For ordinal networks of random series, we present an approach for building the exact form of the adjacency matrix, which in turn is useful for detecting non-random behavior in time series and the existence of missing transitions among ordinal patterns. We find that the average value of a local entropy, estimated from transition probabilities among neighboring nodes of ordinal networks, is more robust against noise addition than the standard permutation entropy. We show that ordinal networks can be used for estimating the Hurst exponent of time series with accuracy comparable with state-of-the-art methods. Finally, we argue that ordinal networks can detect sudden changes in Earth seismic activity caused by large earthquakes.
\end{abstract}

\maketitle

\section{Introduction}
The amount of data researchers currently handle has drastically increased over the past few decades. Not only the data volume has grown, but also the investigated themes and degree of detail are nowadays unprecedented. While large and detailed data sets allow us to probe quantitative questions about topics as diverse as material sciences~\cite{ziletti2018insightful} and art~\cite{sigaki2018history}, this complexity often challenges the methods and techniques available for executing the data analysis. Even the discrimination among simple data such as time series becomes challenging depending on the subject and amount of data involved in the task. Thus, the development of efficient and often interdisciplinary approaches for data analysis represents an important step towards the extraction of hidden and meaningful patterns from complex data sets~\cite{mattmann_computing_2013}.

Among the different approaches that have been proposed, the idea of applying complex network tools for investigating time series has attracted great interest from the scientific community since the seminal works of \citeauthor{nicolis2005dynamical}~\cite{nicolis2005dynamical}, \citeauthor{zhang_complex_2006}~\cite{zhang_complex_2006} and \citeauthor{lacasa_time_2008}~\cite{lacasa_time_2008}, particularly among researchers working on statistical physics~\cite{zou_complex_2019}. These seminal works gave birth to the three main approaches for mapping time series into complex networks: transition networks, proximity networks, and visibility graphs, respectively. In visibility graphs, each data point of a time series is mapped into a vertex of a network, and links are drawn based on a visibility condition between pairs of data points~\cite{lacasa_time_2008,lacasa_description_2010}. Proximity networks, in turn, assume that time series segments play the role of nodes and a similarity measure estimated between every pair of segments defines the network links. Mainly because of its intimate connection with recurrence plots~\cite{zou_complex_2019}, recurrence networks are an important method among the proximity-based approaches. In this approach, time series segments are regarded as a nodes and standard metrics such as Euclidean distance are used to create connections. On the other hand, transition networks consider discrete states over time series partitions as nodes and the connections represent transition probabilities among these states. The ideas underlying transition networks have a long historical precedent that can be traced back to the theory of Markov chains~\cite{schnakenberg1976network}, but its use as a tool for time series analysis is indeed much more recent~\cite{zou_complex_2019}. 

Transition networks of particular interest to this work are the ordinal networks. This method was proposed by \citeauthor{small_complex_2013}~\cite{small_complex_2013} and has a direct inspiration in the work of \citeauthor{bandt_permutation_2002}~\cite{bandt_permutation_2002}, which consists of an embedding method (known as Bandt-Pompe symbolization approach) based on the idea of ordinal or permutation patterns. In ordinal networks, nodes represent possible ordering patterns among elements within partitions of a time series, and links are drawn based on the temporal succession of these patterns. Since its first appearance in the literature, ordinal networks have been mainly used for investigating time series arising from nonlinear dynamical systems (such as those obtained by iterating chaotic maps), while very few works have used these networks for characterizing stochastic processes or real-world time series~\cite{small2018ordinal}. It is also equally surprising that in spite of recently proposed generalizations of this approach (including one dealing with multivariate time series~\cite{zhang2017constructing}), basic properties of ordinal networks obtained from simple univariate stochastic time series (such as a random walk) remain poorly understood. 

Here we help to fill this gap by investigating properties of ordinal networks obtained from periodic, random, and fractional Brownian motion time series. We show that properties of ordinal networks can be used for estimating the Hurst exponent of time series with high precision, outperforming state-of-the-art methods such as detrended fluctuation analysis~\cite{peng_dnadfa_1994,shao2012comparing}. We further demonstrate the usefulness of this algorithm for investigating empirical time series, by showing that ordinal networks are capable of identifying sudden changes in Earth's seismic activity after the occurrence of large earthquakes (mainshocks).

This paper is organized as follows. We review how ordinal networks map time series to complex networks and discuss its general structural constraints in Section~\ref{sec:methods}. Next, in Section~\ref{sec:results}, we investigate properties of ordinal networks emerging from periodic (\ref{sec:results_simple}), fully random (\ref{sec:results_random}), noisy periodic (\ref{sec:results_noise_periodic}), fractional Brownian motion (\ref{sec:results_fBm}), and empirical seismic activity time series (\ref{sec:results_earthquake}). We present our conclusions and final remarks in Section~\ref{sec:conclusions}.

\section{Methods}\label{sec:methods}
As we have mentioned, ordinal networks have a straight connection with the symbolization approach of \citeauthor{bandt2007order}~\cite{bandt2007order} and with the permutation entropy framework of \citeauthor{bandt_permutation_2002}~\cite{bandt_permutation_2002}. To better understand this connection, we start by revisiting such ideas. Within the so-called Bandt-Pompe approach, from a given time series $\{x_t\}_{t = 1, \dots, N}$ of length $N$, we construct $n = N  - d + 1$ overlapping partitions of length $d$ (the embedding dimension), represented by $w_s =  (x_{s}, x_{s + 1}, \dots, x_{s + d - 2}, x_{s + d - 1})$, where $s = 1, \dots, n$ is the partition index. Next, we evaluate the permutation $\pi_s = (r_0, r_1, \dots, r_{d-1})$ of $(0, 1, \dots, d - 1)$ sorting the elements of $w_s$ (in ascending order), that is, the permutations defined by $x_{s - r_{d - 1}} \leqslant x_{s - r_{d - 2}} \leqslant \dots \leqslant x_{s - r_{0}}$. We further assume that $r_i < r_{i - 1}$ if $x_{s - r_{i}} < x_{s - r_{i-1}}$ in case of draws within a partition, keeping the order of occurrence~\cite{cao_detecting_dynamical_2004, rosso_distinguishing_2007}. After these two steps, we obtain a symbolic sequence $\{\pi_s\}_{s = 1, \dots, n}$.

Having obtained all permutations, we thus calculate the relative frequency $p_i(\pi_i)$ of each one of the $d!$ possible permutations $\pi_i$ of the symbols $(0, 1, \dots, d - 1)$
\begin{equation}
    p_i(\pi_i) = \frac{\text{number of partitions of type} \ \pi_i \ \text{in} \ 
    \{\pi_s\}}{n}\,,
\end{equation}
from which we estimate the ordinal probability distribution $P = \{p_i(\pi_i)\}_{i = 1, \dots, d!}$. The permutation entropy is simply the Shannon entropy~\cite{shannon1948mathematical} of the ordinal probability distribution, that is,
\begin{equation}\label{eq:permutation_entropy}
    H = -\sum_{i = 1}^{d!} p_i(\pi_{i})\log p_i(\pi_{i})\,,
\end{equation}
where $\log(\dots)$ stands for the base-$2$ logarithm. The embedding dimension $d>1$ is the only parameter of the approach, and because it defines the number of possible permutations, the condition $d! \ll N$ must hold for a reliable estimate of the distribution $P = \{p_i(\pi_i)\}_{i = 1, \dots, d!}$. The value of $H$ quantifies the randomness in the local ordering patterns of $x_t$: $H\approx\log d!$ indicates that the elements of $x_t$ are locally randomly ordered, while $H\approx0$ implies that the elements of $x_t$ are very likely to appear in a particular order.

Inspired by the Bandt-Pompe approach, \citeauthor{small_complex_2013}~\cite{small_complex_2013} proposed to use the symbolic sequence $\{\pi_s\}_{s = 1, \dots, n}$ for creating the ordinal network, a graph representation of the time series $\{x_t\}_{t = 1, \dots, N}$. The approach consists of considering all possible permutations $\{\pi_i\}_{i=1,\dots,d!}$ as the network nodes, and the connections are drawn between every pair of permutations that occurs in succession within the sequence $\{\pi_s\}_{s = 1, \dots, n}$. The edges are directed according to the temporal succession of permutations (for instance, for the symbolic sequence $\{\pi_1,\pi_2\}$ the link is $\pi_1\to\pi_2$) and are also weighted by the relative frequency in which the succession $(\pi_i,\pi_j)$ occurs in $\{\pi_s\}_{s = 1, \dots, n}$. The elements of the weighted adjacency matrix of this network are
\begin{equation}\label{eq:adj_matrix}
    p_{i, j}\!=\!\frac{\text{number of times} \ \pi_i \ \text{is followed by} \ \pi_{j} \
    {\rm in} \ \{ \pi_s \}}{n - 1}\,,
\end{equation} 
where $i,j=1,2,\dots,d!$ and the denominator $n - 1$ represents the total number of permutation transitions.

\begin{figure}[!ht]
\centering
\includegraphics[width=1\linewidth]{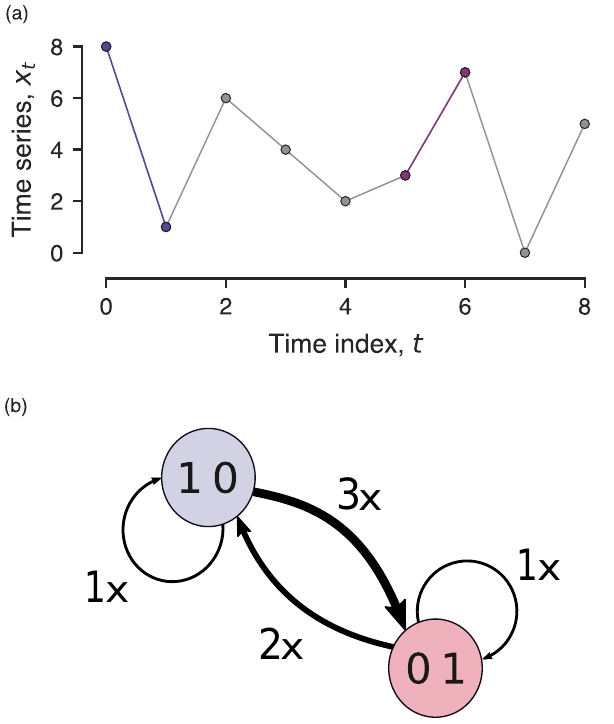}
\caption{Ordinal network approach for mapping time series into networks. (a) Illustration of the time series $x_t = \{8, 1, 6, 4, 2, 3, 7, 0, 5\}$ and the corresponding symbolic sequence $\{(1,0),(0,1),(1,0),(1,0),(0,1),(0,1),(1,0),(0,1)\}$ obtained from the Bandt-Pompe method with embedding dimension $d=2$. (b) Visualization of the ordinal network associated with the time series $x_t$. Nodes `01' (red) and `10' (blue) correspond to the permutations $(0,1)$ and $(1,0)$, respectively. Directed edges are drawn between network nodes based on temporal succession of permutations in the symbolic sequence, and their weights reflect the relative frequency of each possible succession. Self-loops appear when a permutation is followed by itself in the symbolic sequence.}
\label{fig:1}
\end{figure}

To illustrate the ordinal network approach, let us consider a simple time series $x_t = \{8, 1, 6, 4, 2, 3, 7, 0, 5\}$ with $N=9$ elements and embedding dimension $d = 2$ [\autoref{fig:1}(a)]. Our network is composed of two nodes associated with the permutations $\pi_1=(0,1)$ (labeled as `$01$') and $\pi_2=(1,0)$ (labeled as `$10$'). The first partition is $w_1 = (8, 1)$ and corresponds to the permutation $(1,0)$ since the elements of $w_1$ are in descending order. The second partition is $w_2 = (1, 6)$ and corresponds to the permutation $(0,1)$ since the elements of $w_2$ are in ascending order. By repeating this process for all $n=8$ partitions, we obtain the symbolic sequence $\{(1,0),\!(0,1),\!(1,0),\!(1,0),\!(0,1),\!(0,1),\!(1,0),\!(0,1)\}$.~Next, by analyzing all consecutive successions between pairs of permutations, we find that $(0,1)\to(0,1)$, $(0,1)\to(1,0)$, $(1,0)\to(0,1)$, and $(1,0)\to(1,0)$ occur 1, 2, 3, and 1 times, respectively. Thus, the weighted adjacency matrix associated with this particular time series is
\begin{equation*}
\renewcommand\arraystretch{1.2}
\begin{blockarray}{l *{2}{c}}
    \begin{block}{l *{2}{>{$\footnotesize}c<{$}}}
      & (0,1) & (1,0) \\
    \end{block}
    \begin{block}{>{$\footnotesize}l<{$} [*{2}{c}]}
      (0,1)\; & 1/7 & 2/7 \\
      (1,0)\; & 3/7 & 1/7 \\
    \end{block}
  \end{blockarray}\;{\textstyle \raisebox{-5pt}{.}}
\end{equation*}
\autoref{fig:1}(b) shows a visualization of the resulting network. For this illustrative time series, the permutation entropy is $H=1$~bit since $p_1(\pi_1)=1/2$ and $p_2(\pi_2)=1/2$.

By carrying out the parallel with the permutation entropy even further, \citeauthor{mccullough_multiscale_2017}~\cite{mccullough_multiscale_2017} have proposed to estimate another entropic metric based on the transition probabilities (see also \citeauthor{unakafov_conditional_2014}~\cite{unakafov_conditional_2014}). The idea is to calculate a local entropy for node $i$ via
\begin{equation}\label{eq:local_entropy}
    h_{i} = -\sum_{j\in\mathcal{O}_i} p'_{i, j} \log p'_{i, j}\,,
\end{equation}
where $p'_{i, j} = p_{i,j} / \sum_{k\in\mathcal{O}_i} p_{i,k}$ is the renormalized transition probability of leaving node $i$ to node $j$ and $\mathcal{O}_i$ stands for the outgoing neighbourhood of node $i$ (all edges that leave node $i$). The value of $h_{i}$ thus quantifies the degree of randomness associated with transitions starting from permutation $\pi_i$. For instance, $h_{i} = \log \vert\mathcal{O}_i\vert$ ($\vert\dots \vert$ stands for set cardinality, that is, the number of possible outgoing neighbours of $i$) if all possible transitions leaving $\pi_i$ are equiprobable, whereas $h_{i}=0$ when there exists only one transition leaving $\pi_i$. For the network in~\autoref{fig:1}, $h_1=h(\text{`01'})=0.918$ and $h_2=h(\text{`10'})=0.811$. 

From the value of $h_{i}$, \citeauthor{mccullough_multiscale_2017}~\cite{mccullough_multiscale_2017} have proposed to calculate the global node entropy
\begin{equation}\label{eq:global_entropy}
    H_{\text{GN}} = \sum_{i=1}^{d!} p'_{i}\, h_{i}\,,
\end{equation}
where $p'_{i} = \sum_{j\in\mathcal{I}_i} p_{j,i}$ is the probability of arriving at node $i$ from its incoming neighbourhood $\mathcal{I}_i$ (in-strength of node $i$). It is worth noticing that $p'_{i}\approx p_i(\pi_i)$ for long time series, and thus, the permutation entropy is the Shannon entropy associated with the in-strengths of ordinal network nodes. \autoref{eq:global_entropy} represents a weighted mean of the local node entropy, and quantifies the global degree of randomness over all (first order) permutation transitions occurring in a time series $x_t$. More regular time series have smaller values of $H_{\text{GN}}$ than time series associated with random processes. For the example shown in~\autoref{fig:1}, we find that $H_{\text{GN}}=0.872$. \citeauthor{mccullough_multiscale_2017}~\cite{mccullough_multiscale_2017} have further proposed to ignore the auto-loops when estimating the global node entropy ($H_{\text{GN}}$). As they have argued, auto-loops may become too intense when dealing with over-sampled time series, which in turn may bias the values of $H_{\text{GN}}$. In our work, we have opted to consider the auto-loops because of the nature of the time series we shall investigate. To be more precise in terminology, the entropy definition of Eq.~\ref{eq:global_entropy} is usually called conditional permutation entropy~\cite{unakafov_conditional_2014}. It is also worth mentioning that our entropy definition and other topological properties of ordinal networks are different from the concept of simplicial complexes of graphs~\cite{jonsson2008simplicial}, where algebraic topology methods are used for defining topological information measures~\cite{baudot2019topological}, such as the entropy of a topological level used by \citeauthor{andjelkovic2015hidden}~\cite{andjelkovic2015hidden} for investigating traffic jamming.

In addition to global node entropy $H_{\text{GN}}$, the study of time series based on ordinal networks explores the myriad of network metrics available for characterizing complex networks. However, important features of ordinal networks are naturally inherited and limited by the symbolic sequences that give rise to networks. Some of these properties of ordinal networks have already been implicitly discussed in previous works~\cite{small_complex_2013,mccullough_time_lagged_2015,sun_characterizing_2014}, but they still lack attention. An important limitation is related to the maximum number of connections for a node. This constraint is not readily apparent when $d = 2$, because this case displays all possible connections [\autoref{fig:1}(b)]. On the other hand, there exist six possible permutations (that is, $d!$) and thus six nodes in ordinal networks when $d = 3$, but each node connects with only three other nodes. This restriction arises from the fact that a permutation $\pi_s$ ordering a partition $w_s$ can only be followed by three ordinal patterns associated with the partition $w_{s+1}$. 

To illustrate such constraints, let us consider a partition $w_1 = (x_1, x_2, x_3)$ having the permutation $\pi_1 = (0,1,2)$, that is, the ordering of its elements is such that $x_1 < x_2 < x_3$. Suppose now that the next partition is $w_2 = (x_2, x_3, x_4)$, where $x_2$ and $x_3$ are the same elements contained in $w_1$. Because $\pi_1 = (0,1,2)$, we know that the condition $x_2 < x_3$ should hold in $w_2$, and therefore, the number $0$ should precede $1$ in the permutation associated with $w_2$. Among all possible permutations for $d=3$, $\{(0,\! 1,\! 2), (0,\! 2,\! 1), (1,\! 0,\! 2), (1,\! 2,\! 0), (2,\! 0,\! 1), (2,\! 1,\! 0)\}$, only $(0,\!1,\!2), (0,\!2,\!1)$ and $(2,\!0,\!1)$ satisfy the previous condition. Thus, there are only three permutations $\pi_2$ that can appear after $\pi_1$, which in turn limit the number of outgoing connections of node $(0,1,2)$ to three. By using the same argument, we can verify that the number of incoming connections is also limited to three and that these constraints hold for all nodes. 

These ideas generalize to all values of embedding dimension $d$, so that all nodes in a permutation network have in-degree and out-degree limited to numbers between $0$ and $d$. Consequently, ordinal networks cannot have more than $d \times (d!)$ edges. This limitation results from the fact that the ordering of the elements in a partition $w_s$ is partially carried out to the next partition $w_{s+1}$, since the elements of $w_{s+1}$ comprise $d - 1$ elements of $w_s$. Furthermore, the smaller the embedding dimension, the fewer are the elements shared among the time series partitions, and thus, the series past rapidly becomes unimportant for determining future permutations. 

Another intriguing consequence of the constraints associated with transitions among permutations is related to self-edges in the network. By analyzing all transitions, we conclude that self-loops can only exist for two particular nodes of ordinal networks, regardless of the embedding dimension $d$. These two nodes are associated with only ascending or only descending permutations, that is, permutations related to partitions in which the elements are all successively increasing or decreasing. For instance, these nodes correspond to the permutations $(0,1,2)$ and $(2,1,0)$ in the case of $d=3$, and $(0,1,2,3)$ and $(3,2,1,0)$ for $d=4$. 

It is worth noticing that other generalized algorithms for building ordinal networks from time series have been proposed and applied. These different numerical recipes include the use of non-overlapping partitions~\cite{small_complex_2013, masoller_quantifying_2015}, partitions with time-lagged elements~\cite{mccullough_time_lagged_2015}, and the inclusion of amplitude information about the time series elements~\cite{sun_characterizing_2014}. Naturally, the constraints we have discussed here do not hold for these generalized algorithms. For instance, a given permutation $\pi_s$ can be followed by any possible permutation in the case of non-overlapping partitions since consecutive partitions do not share any time series elements. For time-lagged elements, similar restrictions among permutations emerge for high-order transitions. These generalized algorithms are also interesting and may deserve further investigation; here, we have focused on ordinal networks directly inspired by the seminal works of  \citeauthor{bandt2007order}~\cite{bandt2007order} and  \citeauthor{bandt_permutation_2002}~\cite{bandt_permutation_2002}, as illustrated in \autoref{fig:1}.

\section{Results}\label{sec:results}
\subsection{Ordinal networks of simple time series}\label{sec:results_simple}
We start our empirical investigation by analyzing the structure of ordinal networks arising from elementary time series. Perhaps the simplest time series to consider is a monotonic (increasing or decreasing) series. In this case, regardless of the embedding dimension $d$, the ordinal network is composed of a single node (representing the solely increasing or decreasing permutation) with only one auto-loop (meaning that the permutation is always followed by itself). Therefore, network metrics for monotonic time series are all trivial. 

\begin{figure*}[!ht]
\centering
\includegraphics[width=0.85\linewidth]{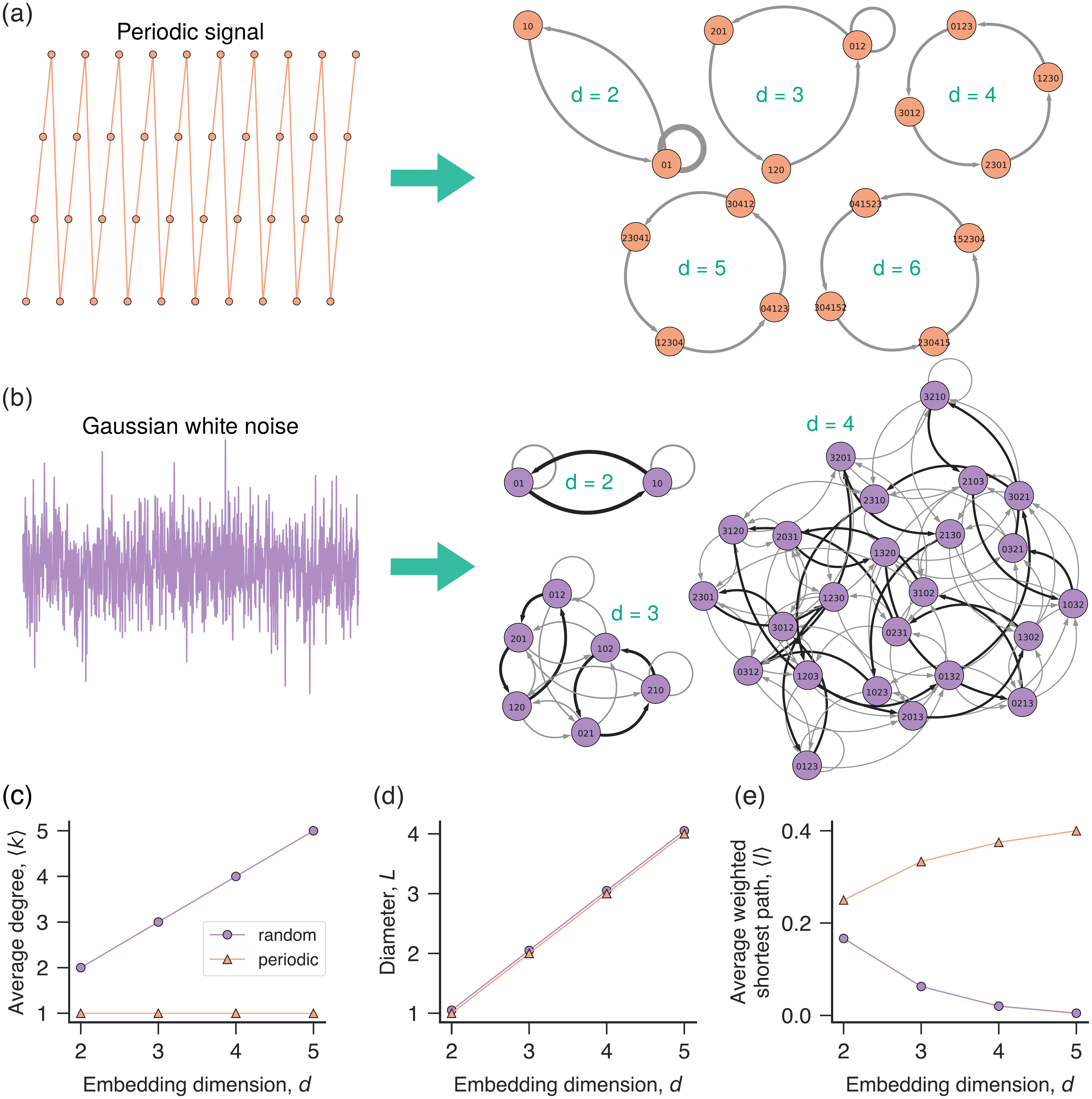}
\caption{Ordinal networks of periodic and random time series. (a) Illustration of the mapping of a periodic signal into ordinal networks with different embedding dimensions $d$ (indicated within the panel). (b) Mapping of Gaussian white noise into ordinal networks with different embedding dimensions $d$ (shown within the panel). Black edges indicate the transitions that are twice as likely to occur. (c) Average degree $\langle k \rangle$, (d) diameter $L$, and (e) average weighted shortest path $\langle l \rangle$ as a function of the embedding dimension $d$ for ordinal networks emerging from random (circles) and periodic (triangles) signals. The periodic signals have period $T$ matching the embedding dimension ($d = T$). The results for the three previous properties hold for sufficiently long time series, that is, when a reliable estimate of all permutation transitions is available.}
\label{fig:2}
\end{figure*}

Periodic series are another example of simple yet more interesting signals. Ordinal networks for simple periodic series form cyclic structures, where the arrangement of nodes and edges allude to the series itself. If the number of data points within a period of a time series is $T$ and the embedding dimension is $d = T$, the ordinal network is a cyclic graph (without auto-loops) with $T$ nodes (number of different permutations) since the time series repeats itself after $T$ data points. Furthermore, this cyclic structure does not change if we consider larger embedding dimensions ($d > T$), only the permutation symbols associated with the network nodes are modified [\autoref{fig:2}(a)]. Because of this behavior, ordinal networks of periodic signals display a constant average degree [for in and out connections, \autoref{fig:2}(c)] and a diameter that grows linearly with the embedding dimension [\autoref{fig:2}(d)]. We can also show that the average weighted distance from one node to all others is $\frac{1}{d}\sum_{i = 1}^{d -1} \left( \frac{i}{d} \right)$, which is independent of the node we choose due to graph symmetry. Thus, the average weighted shortest path for the whole network is also
\begin{equation}\label{eq:avg_distance_periodic}
    \langle l \rangle_{\rm per} = \frac{1}{d}\sum_{i = 1}^{d -1} \left( \frac{i}{d} \right) = \frac{d-1}{2 d}\,,
\end{equation}
where the summation represents the average weighted distance from a particular node to all others, and the factor $1/d$ accounts for the average over all nodes in the network [\autoref{fig:2}(e)]. This expression is valid when all existing edges have the same weight, which holds for long time series or series composed of an integer number of periods. 

\subsection{Random ordinal networks}\label{sec:results_random}
We have further investigated ordinal networks emerging from random signals. Differently from monotonic and periodic signals, ordinal networks emerging from random series (henceforward called random ordinal networks) are expected to display all possible connections since all permutations are equally likely to occur in random signals that are sufficiently long. \autoref{fig:2}(b) shows examples of ordinal networks for different embedding dimensions emerging from Gaussian white noise. We notice that these networks are composed of $d!$ nodes and each node has $d$ outgoing and $d$ incoming links. Because of these properties, the average degree and the diameter of random ordinal networks increase linearly with the embedding dimension $d$ [\autoref{fig:2}(c) and (d)].

An intriguing aspect of these random ordinal networks is that edge weights are not all the same, albeit all permutations are equiprobable in random time series. Indeed, one of the $d$ outward edges of all nodes has twice the weight of the remaining $d - 1$ edges [black links in \autoref{fig:2}(b)], a result that holds for long sequences of random numbers drawn from any continuous probability distribution. To illustrate how this happens, let us consider that the first partition with $d=3$ associated with a random series is $w_1 = (x_1, x_2, x_3)$ and the corresponding permutation is $\pi_1 = (0,1,2)$ (that is, $x_1<x_2<x_3$). The next partition is $w_2 = (x_2, x_3, x_4)$ and the new element $x_4$ fits one of the following conditions: {\it i)} $x_4<x_1<x_2<x_3$, {\it ii)} $x_1<x_4<x_2<x_3$, {\it iii)} $x_1<x_2<x_4<x_3$, and {\it iv)} $x_1<x_2<x_3<x_4$. Conditions {\it i)} and {\it ii)} yield $\pi_2=(2,0,1)$, while {\it iii)} results in $\pi_2=(0,2,1)$ and {\it iv)} in $\pi_2=(0,1,2)$. Thus, there are two possibilities of finding $\pi_2=(2,0,1)$ which makes the transition  $(0,1,2)\to(2,0,1)$ twice as likely than $(0,1,2)\to(0,2,1)$ or $(0,1,2)\to(0,1,2)$. It is also worth noticing that if we draw a large number of samples $(x_1, x_2, x_3)$ so that $x_1<x_2<x_3$, the average values of $x_1$, $x_2$, and $x_3$ converge to the quartiles of the probability distribution of the time series, and therefore, the conditions {\it i)-iv)} are equiprobable. The same idea holds for all other first-order transitions when $d=3$ and for any other embedding dimension. 

A general rule of thumb for determining the permutation $\pi_{s+1}$ that is more likely to follow $\pi_{s}$ is picking the $\pi_{s+1}$ in which the symbol equal to `$d-1$' fits the position of the symbol `$0$' in $\pi_{s}$. For instance, if $\pi_{s}=(3,2,1,0)$, $\pi_{s+1}^{*}=(2,1,0,3)$ is twice as likely than $\pi_{s+1}\in\{(3,2,1,0),(2,1,3,0),(2,3,1,0)\}$ because the symbol `3' in $\pi_{s+1}^{*}$ is located in the same position symbol `0' is placed in $\pi_s$. This result allows us to build the weighted adjacency matrix of random ordinal networks for any embedding dimension $d$. To do so, we start with a network having $d!$ nodes (each one representing a particular permutation $\pi_i$), and draw a directed connection with unitary weight between all pairs of nodes $\pi_i$ and $\pi_j$ for which the transition $\pi_i\to\pi_j$ is possible. Next, we update the weights of all more probable connections from one to two. Finally, the elements of the resulting adjacency matrix $p_{i,j}$ are divided by the factor $(d + 1)!$, which represents the sum of all unitary weights [$d! (d-1)$] plus twice the number of edges with double weights [$d!(2)$]. By using this weighted adjacency matrix, we can numerically evaluate any network metric of random ordinal networks with arbitrary embedding dimension $d$. \autoref{fig:2}(e) shows the average weighted shortest path as a function of $d$ estimated from these theoretical networks, where we observe that this measure approaches zero for large values of $d$ since edge weights decrease with the increase of $d$.

Another interesting property of random ordinal networks that we can analytically estimate is the global node entropy (Eq.~\ref{eq:global_entropy}). To do so, we first notice that the local entropy for node $i$ is
\begin{equation}
\begin{split}
    h_i & = - \tilde{p}'_{i, j} \log \tilde{p}'_{i, j} - \sum_{j=1}^{d-1} p'_{i, j} \log p'_{i, j}\\
        & = -\frac{2}{d+1} \log \left( \frac{2}{d+1} \right) - \left( \frac{d-1}{d+1} \right) \log \left( \frac{1}{d+1} \right)\,,
\end{split}
\end{equation}
where $p'_{i, j} = 1/(d+1)$ are the renormalized transition probabilities from node $i$ to node $j$ that are equiprobable, and $\tilde{p}'_{i, j} = 2 p'_{i, j}$ represents the renormalized transition probability that is twice as likely to occur. By plugging this result into Eq.~\ref{eq:global_entropy} with $p_i'=1/d!$ (equiprobable permutations), we find that the global node entropy for uncorrelated sequences of random numbers is
\begin{equation}\label{eq:global_entropy_random}
    H_{\text{GN}}^{\text{(rand)}} = \log(d+1) - (\log 4)/(d+1)\,.
\end{equation}
We note that the value of $H_{\text{GN}}^{\text{(rand)}}$ is always smaller than the one obtained when all transitions are equiprobable (for this case, $H_{\text{GN}}^{\text{(equi)}}=\log d$), and only when $d\to\infty$ we have $H_{\text{GN}}^{\text{(rand)}}\to H_{\text{GN}}^{\text{(equi)}}$. Thus, the value of $H_{\text{GN}}^{\text{(rand)}}$ does not represent the maximum entropy for a given $d$, in a sense that there exist time series for which $H_{\text{GN}}>H_{\text{GN}}^{\text{(rand)}}$.

\begin{figure*}[!ht]
\centering
\includegraphics[width=1\linewidth]{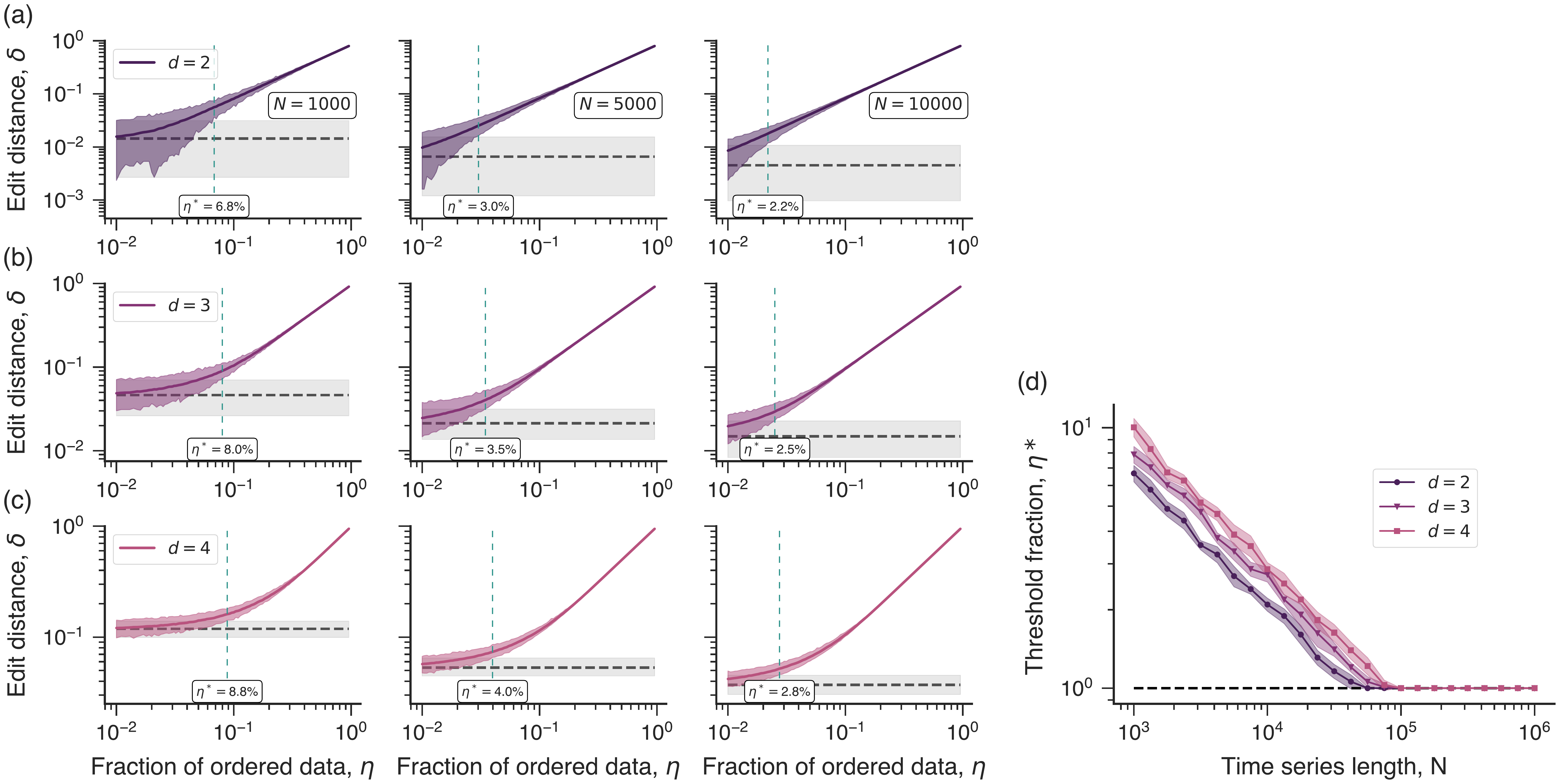}
\caption{Detecting non-random behavior with random ordinal networks. Panels (a)-(c) show the average values of the edit distance $\delta$ (colored lines) between ordinal networks of partially ordered white noise ($\eta$ is the fraction of ordered elements) and the theoretical random ordinal networks for embedding dimensions $d=2,3,$ and $4$, respectively. In each of these panels, results of the first column are obtained from ensembles of $1000$ time series with $N=1000$ data points each, while the second and third columns show the results from ensembles of $1000$ series with $N=5000$ and $10000$ elements, respectively. The colored shaded regions stand for $95\%$ confidence bands, and the gray shaded regions are $95\%$ random confidence bands for $\delta$ calculated from finite random series. Vertical dashed lines indicate the threshold fraction $\eta^{*}$ of ordered elements at which a partially ordered random series becomes distinguishable from a totally random series of the same length. (d) Threshold fraction $\eta^{*}$ as a function of the time series length $N$ for different embedding $d\in\{2,3,4\}$. The markers represent average value of $\eta^{*}$ estimated from $10$ replicas, the shaded regions are one standard deviation bands, and the dashed line indicates the minimum fraction of ordered elements used in our numerical experiments ($1\%$).}
\label{fig:3}
\end{figure*}

An interesting application for these theoretical random networks is to detect non-random behavior in empirical time series. One option is to directly compare empirical ordinal networks with their random theoretical counterparts via a proper graph distance measure~\cite{donnat2018tracking, wills2019metrics}. To illustrate this possibility, we have investigated whether partially ordered Gaussian white noise can be distinguishable from pure noise by estimating the edit distance $\delta$~\cite{wills2019metrics} among their ordinal networks. For a weighted and directed graph, this simple metric represents the amount of edge weight (strength) that needs to be reallocated so that the two graphs become equal. We have generated an ensemble of white noise series, where a given fraction $\eta$ of consecutive elements are placed in ascending order. Next, we map these time series into ordinal networks and estimate the average values of the edit distance $\delta$ to the theoretical random networks as a function of the fraction of ordered data $\eta$. Figures~\ref{fig:3}(a)-(c) show the behavior of the average value of $\delta$ as a function of $\eta$ as well as the 95\% confidence region for embedding dimensions $d\in\{2,3,4\}$ and for series with $N\in\{1000,5000,10000\}$ elements.

As expected, we observe that the values of $\delta$ increase with $\eta$ in all cases. We have also estimated the average value of the edit distance between empirical ordinal networks of white noise series (with the same lengths) and the theoretical random networks for constructing random confidence bands. Values of $\delta$ outside this random confidence band represent a reliable indicator that an empirical time series displays deviations from pure random behavior. Moreover, these confidence regions also allow us to identify the threshold fractions $\eta^{*}$ of order from which the edit distance $\delta$ is capable of accurately detecting this anomaly. Figures~\ref{fig:3}(a)-(c) show that this approach is capable of detecting the ordering in time series with 1000 elements for $\eta^{*}\approx6.8\%$ when $d=2$, for $\eta^{*}\approx8.0\%$ when $d=3$, and for $\eta^{*}\approx8.8\%$ when $d=4$. These figures also indicate that the values of $\eta^{*}$ decrease with the increase of the time series length $N$. However, we observe that $d=2$ displays the smallest values of $\eta^{*}$ regardless of $N$. 

We have further estimated the average values of $\eta^{*}$ over $10$ realizations of the detection procedure as a function of $N$. \autoref{fig:3}(d) shows that $\eta^{*}$ decreases exponentially with $N$ for $d\in\{2,3,4\}$ and that for very long series ($N>10^5$) all values of $d$ are equally efficient in detecting the minimal fraction of ordered elements used in our numerical experiments ($\eta=1\%$). We observe that $d=2$ produces the smallest threshold fractions, but the proximity of the one standard deviation confidence bands between $d=2$ and $d=3$ indicates that the difference between these cases is tight. This happens because we create disproportionately more ascending ordinal patterns when we partially sort a series, and since ordinal networks with $d=2$ have only four links (and two nodes), this imbalance is more easily detected than when considering ordinal networks with higher values of $d$. This situation is likely to change if the anomalous pattern becomes more complex, so that ordinal networks with higher embedding dimensions may perform better at detecting the anomaly in such cases. However, this example illustrates that there will always be a trade-off between embedding dimension and time series length.

\begin{figure}[!t]
\centering
\includegraphics[width=0.86\linewidth]{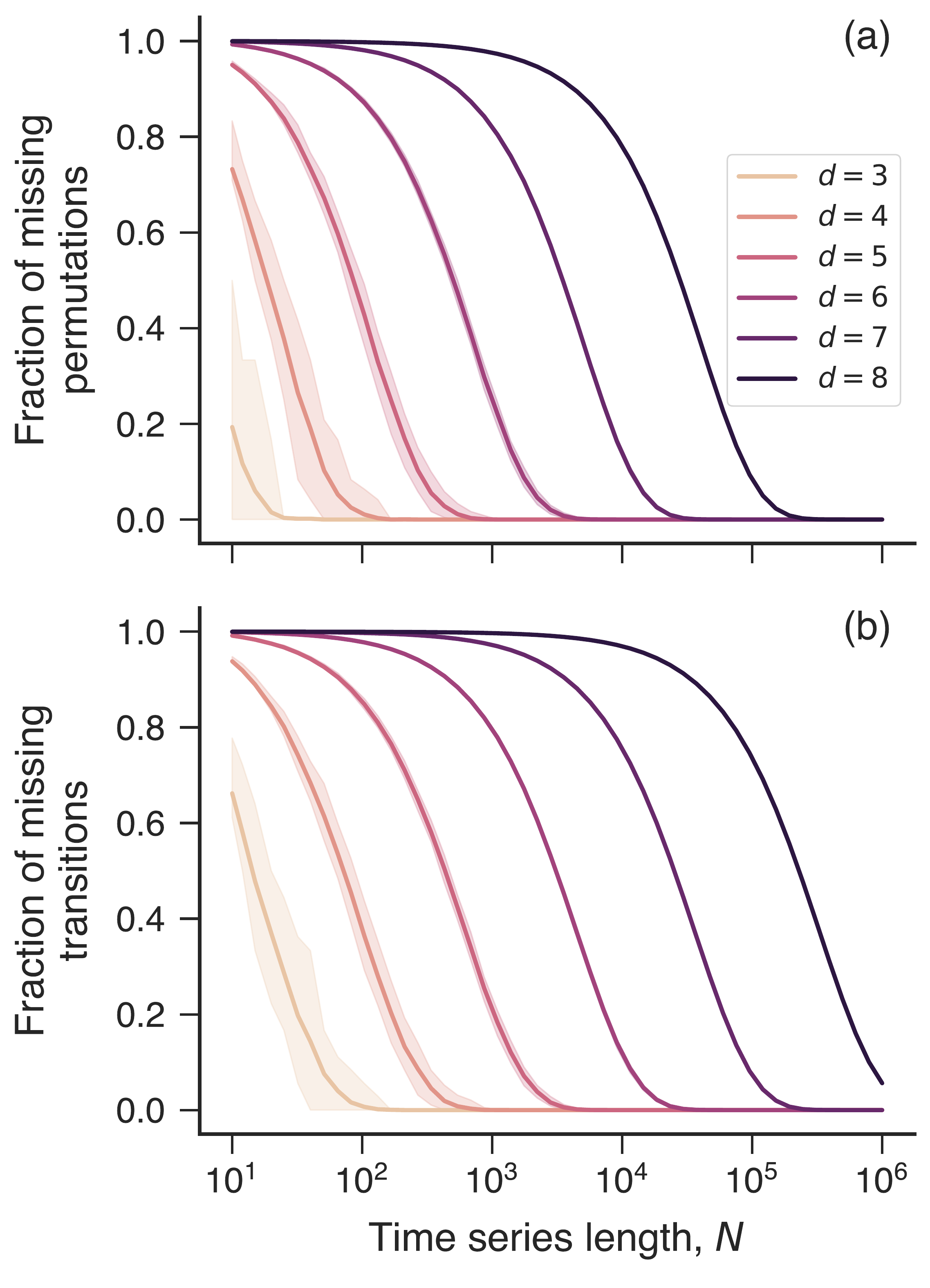}
\caption{Missing ordinal patterns and missing permutation transitions in random series. (a) Fraction of missing permutations and (b) fraction of missing permutation transitions in Gaussian white noise series of unitary variance as a function of the time series length $N$ for different values of embedding dimensions $d\in\{3, 4, 5, 6, 7, 8\}$ (indicated by the legend). In both plots, the colored curves represent average values over 100 realizations of white noise series for each embedding dimension, and the shaded regions are 95\% confidence intervals.}    
\label{fig:4}
\end{figure}

By knowing the exact form of random ordinal networks, we can also estimate the fraction of missing transitions in empirical time series. This analysis is somewhat similar to the one of missing permutations or forbidden patterns introduced by \citeauthor{amigo2006order}~\cite{amigo2006order,amigo_true_2007,amigo2008combinatorial} and explored by several works~\cite{mccullough2016countingI,sakellariou2016countingII,olivares2019revisiting}. These works have observed that some ordinal patterns cannot occur in chaotic systems [for instance, the permutation $(2,1,0)$ never appears in logistic map under fully developed chaos] and that even stochastic processes may present missing ordinal patterns depending on the time series length and the choice of embedding dimension. However, the number of missing patterns in random processes decays as the time series become longer, and because of that, are often called ``false forbidden patterns.'' \autoref{fig:4}(a) illustrates how the fraction of missing permutations in Gaussian white noises decreases with the increase of the time series length for different embedding dimensions $d\in\{3, 4, 5, 6, 7, 8\}$. To extend these results, we test whether noisy series also present ``false forbidden transitions''. To do so, we evaluate ordinal networks from Gaussian white series of length $N$ and compare their number of transition links with those in the exact form of the random ordinal network. \autoref{fig:4}(b) shows these values for different embedding dimensions, where we observe that white noise series also present ``false missing transitions'' depending on the values of $N$ and $d$. We note that the shape of these curves resembles those observed for the fraction of missing permutations; however, the number of missing transitions decreases more slowly with $N$ than the number of missing permutations. It is also worth mentioning that the edit distance calculated between these empirical ordinal networks and the exact form of random ordinal networks quantifies not only the existence of missing transitions, but the differences in their occurrence frequencies in a time series.

\begin{figure*}[!ht]
\centering
\includegraphics[width=1\linewidth]{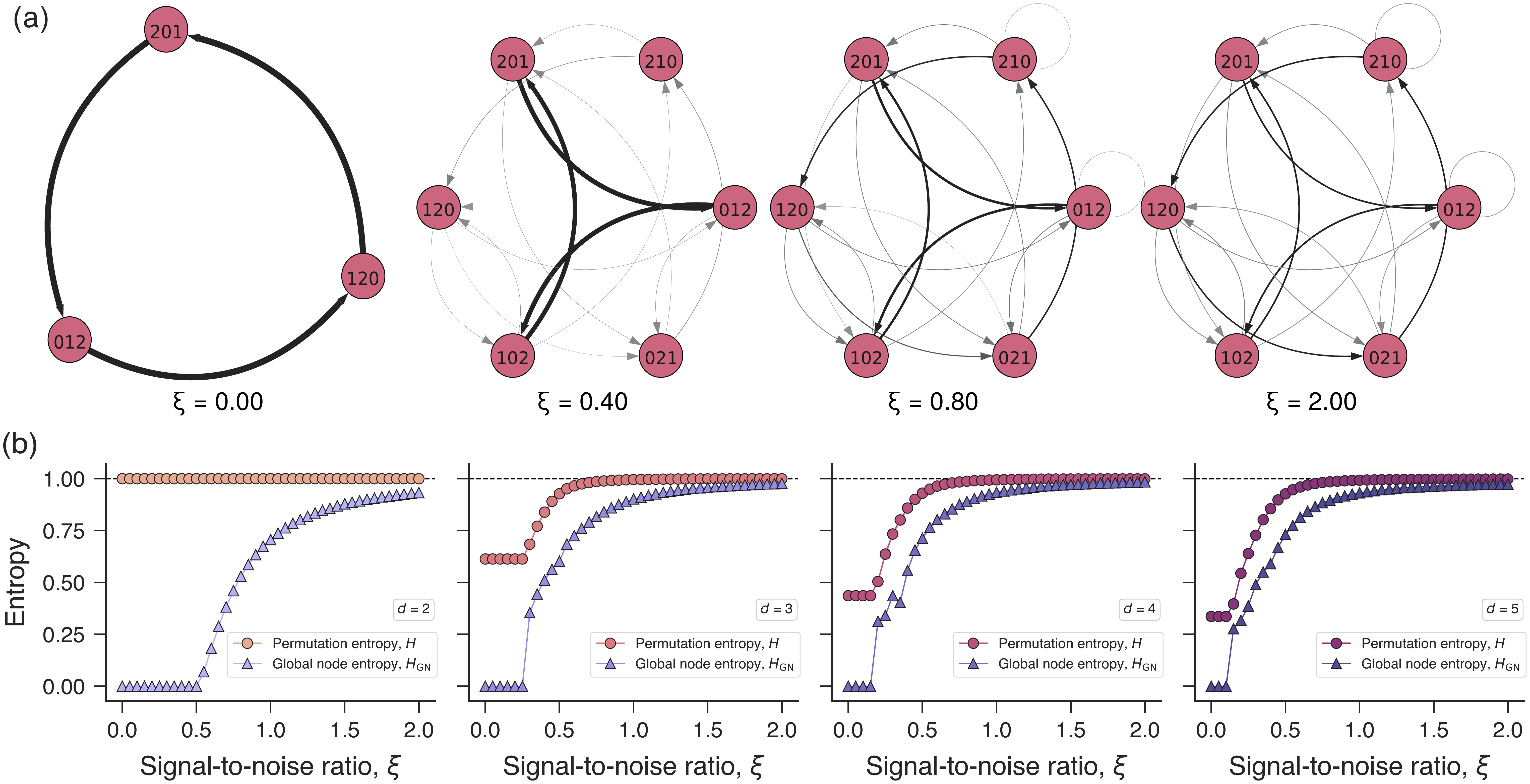}
\caption{Robustness to noise addition of permutation entropy and global node entropy. (a) Visualizations of ordinal networks related to noisy periodic signals for $d=3$ and different signal-to-noise ratios $\xi$ (shown within the panel). (b) Average values of the permutation entropy $H$ (circles) and global node entropy $H_{\text{GN}}$ (triangles) obtained from an ensemble of $100$ realizations of noisy periodic signals as a function of $\xi$ for $d=2,3,4$ and $5$. Each time series has $10^{4}$ elements and the period $T$ is set equal to the embedding dimension ($T =d$). The permutation entropy is normalized by its maximum value and the global node entropy by the value of the random ordinal network (Eq.~\ref{eq:global_entropy_random}). The tiny shaded regions correspond to one standard deviation band, and dashed lines are the values for random ordinal networks.}
\label{fig:5}
\end{figure*}

\subsection{Ordinal networks of noisy periodic time series}\label{sec:results_noise_periodic}
It is also interesting to investigate how ordinal networks of periodic signals transform into random ordinal networks as we add noise to such time series. To do so, we generate sawtooth-like signals of length $10^{4}$ and period $T$ of the form $x_t = \{0, 1, 0, 1, \dots\}$ ($T=2$), $x_t = \{0, 1/2, 1, 0, 1/2, 1, \dots\}$ ($T = 3$), and so on. Next, all elements of these time series are incremented with a uniformly distributed noise over the interval $[-\xi,\xi]$, where $\xi$ is a parameter controlling the signal-to-noise ratio. \autoref{fig:5}(a) shows examples of ordinal networks obtained from these noisy periodic signals for $\xi\in\{0,0.4,0.8,2\}$ and $d=T=3$. As expected, the ring-like structure of the periodic ordinal network ($\xi=0$) approaches a random ordinal network as we increase the value of $\xi$. We notice that this gradual process starts with the emergence of all possible network nodes (which happens for small values of $\xi$), followed by the fading of the three initial links of the network for $\xi=0$ and the enhancement of the other links. We have also verified that edit distance calculated between ordinal networks of noisy periodic signals and the exact form of random ordinal networks approaches zero as the signal-to-noise ratio increases.

These noisy periodic time series also allow us to test the robustness to noise addition of the global node entropy $H_{\text{GN}}$ (Eq.~\ref{eq:global_entropy}) in comparison with the standard permutation entropy $H$ (Eq.~\ref{eq:permutation_entropy}). For that purpose, we generate an ensemble of $100$ sawtooth-like signals of length $10^{4}$ for each value of $\xi\in\{0,0.05,0.1,\dots,2\}$. Next, we estimate the values of $H_{\text{GN}}$ and $H$ for each time series and the average values of these quantities for all $\xi$ values. We have further normalized these quantities dividing the values of $H_{\text{GN}}$ by $H_{\text{GN}}^{\text{(rand)}}$ (Eq.~\ref{eq:global_entropy_random}) and $H$ by $\log d!$. \autoref{fig:5}(b) shows the mean of the normalized values of $H_{\text{GN}}$ and $H$ as a function of $\xi$ for $d=2,3,4$ and $5$. We notice that the values of $H_{\text{GN}}$ are more robust than $H$ against noise addition, and therefore, more efficient for distinguishing among time series with different values of $\xi$. This feature is more evident for $d=2$ (and $T=2$) because in this case $x_t=\{0,1,0,1,\dots\}$, and so the two ordinal patterns are equiprobable even when $\xi<0.5$, making the normalized value of $H$ always equal to 1. Conversely, the global node entropy is zero for $\xi<0.5$ and starts to increase for values of $\xi$ greater than $0.5$. For larger embedding dimensions, we observe that the permutation entropy $H$ approaches $1$ much faster than $H_{\text{GN}}$. For instance, the values of $H$ are unable to distinguish these noisy periodic series if $\xi>1$, whereas the values of $H_{\text{GN}}$ are still able to differentiate them.

\subsection{Ordinal networks of fractional Gaussian noise and fractional Brownian motion}\label{sec:results_fBm}

We have also investigated ordinal networks emerging from time series of fractional Gaussian noise and fractional Brownian motion~\cite{Mandelbrot1982,mandelbrot_fractional_1968}. Fractional Brownian motion is a stochastic process characterized by a parameter $h\in (0,1)$ (the Hurst exponent) that exhibits long-range correlations. The increments of a fractional Brownian motion are Gaussian distributed with zero mean, stationary, and usually called fractional Gaussian noise. The Hurst exponent $h$ controls the roughness/raggedness of these stochastic processes. For $h<1/2$, a fractional Gaussian noise is anti-persistent, meaning (roughly speaking) that positive values are followed by negative values (or vice versa) more frequently than by chance. On the other hand, a fractional Gaussian noise is persistent if $h>1/2$, meaning that positive values are followed by positive values, and negative values are followed by negative values more frequently than by chance; uncorrelated (white) noise corresponds to $h\to1/2$. For fractional Brownian motion, the larger the value of $h$, the smoother is the generated time series. 

\begin{figure*}[!ht]
\centering
\includegraphics[width=1\linewidth]{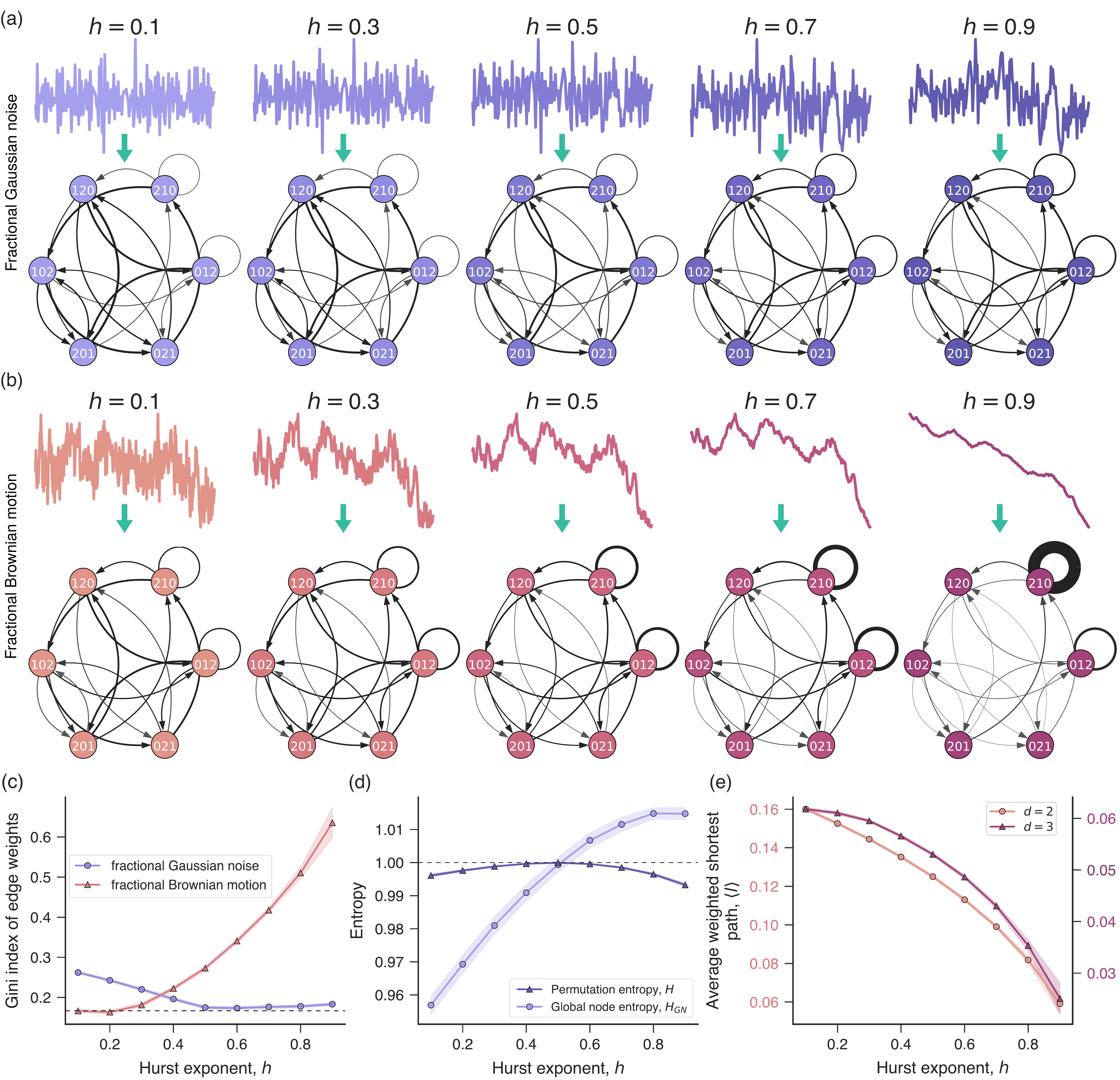}
\caption{Ordinal networks of fractional Gaussian noises and fractional Brownian motions. (a) Illustration of the mapping of fractional Gaussian noise samples with different Hurst exponents ($h$, shown within the panel) into ordinal networks. (b) Illustration of the mapping of fractional Brownian motion samples with different Hurst exponents ($h$, shown within the panel) into ordinal networks. In the previous panels, the thickness of the links are proportional to the edge weights and all time series are generated with the same starting seed. (c) Dependence of the Gini index associated with the edge weights distribution on the Hurst exponent $h$ for the fractional Gaussian noise (circles) and fractional Brownian motion (triangles). (d) Permutation entropy $H$ (triangles) and global node entropy $H_{\text{GN}}$ (circles) with $d = 3$ as a function of the Hurst exponent $h$ for the fractional Gaussian noise. The permutation entropy is normalized by its maximum value and the global node entropy by the value of the random ordinal network (Eq.~\ref{eq:global_entropy_random}). (e) Average weighted shortest path $\langle l \rangle$ as a function of the Hurst exponent $h$ with $d=2$ (circles) and $d=3$ (triangles) for the fractional Brownian motion. In panels (c)-(e), all curves represent average values estimated from an ensemble of $1000$ series for each Hurst exponent, and the shaded areas represent one standard deviation band. The dashed lines in panels (c) and (d) represent the metric values obtained from random ordinal networks.\vspace*{2cm}}
\label{fig:6}
\end{figure*}

\autoref{fig:6}(a) and (b) show examples of fractional Gaussian noises and fractional Brownian motions for different values of $h$ generated with the procedure of Hosking~\cite{Hosking1984}. These figures also depict visualizations of the corresponding ordinal networks for $d=3$ obtained for each particular realization of both processes. We observe that all possible nodes are present in these networks and that every node makes all allowed connections. Therefore, these networks would be identical without edges weights. Indeed, a visual inspection of the weight patterns of these ordinal networks already informs us about particularities of each time series. For fractional Gaussian noises, we observe an uneven weight distribution for small values of the Hurst exponent $h$, while fractional Brownian motions display more balanced weights for low values of $h$. In the case of fractional Brownian motions, it is worth noticing that the auto-loop weight associated with the permutation $(2,1,0)$ becomes quite intense for large values of $h$, reflecting the downward trends of those particular realizations of the stochastic process.

One possibility for quantifying inequality in edge weights is to calculate the Gini index~\cite{gini1921measurement}. This coefficient is widely used in several disciplines (especially in economics) and represents a measure of statistical dispersion of probability distributions. Values of Gini index close to zero show that the weights are equally distributed, while values close to 1 indicate a sharp inequality in the weight distribution. We have estimated the Gini index from an ensemble of ordinal networks associated with fractional Gaussian noise and fractional Brownian motion with different Hurst exponents (100 realizations for each $h \in \{0.1, 0.2, \dots, 0.9\}$). The average values of the Gini coefficient are shown in \autoref{fig:6}(c), where the results confirm our previous visual analysis. For fractional Gaussian noise, we observe that the Gini index decreases as the values of $h$ rise up to $h\approx0.5$, from where it displays a plateau whose value is very close to the Gini index of a random ordinal network. Conversely, the Gini coefficient systematically increases with $h$ for the fractional Brownian motion, reflecting the rise in the persistent behavior. 

\begin{figure*}[!ht]
\centering
\includegraphics[width=1\linewidth]{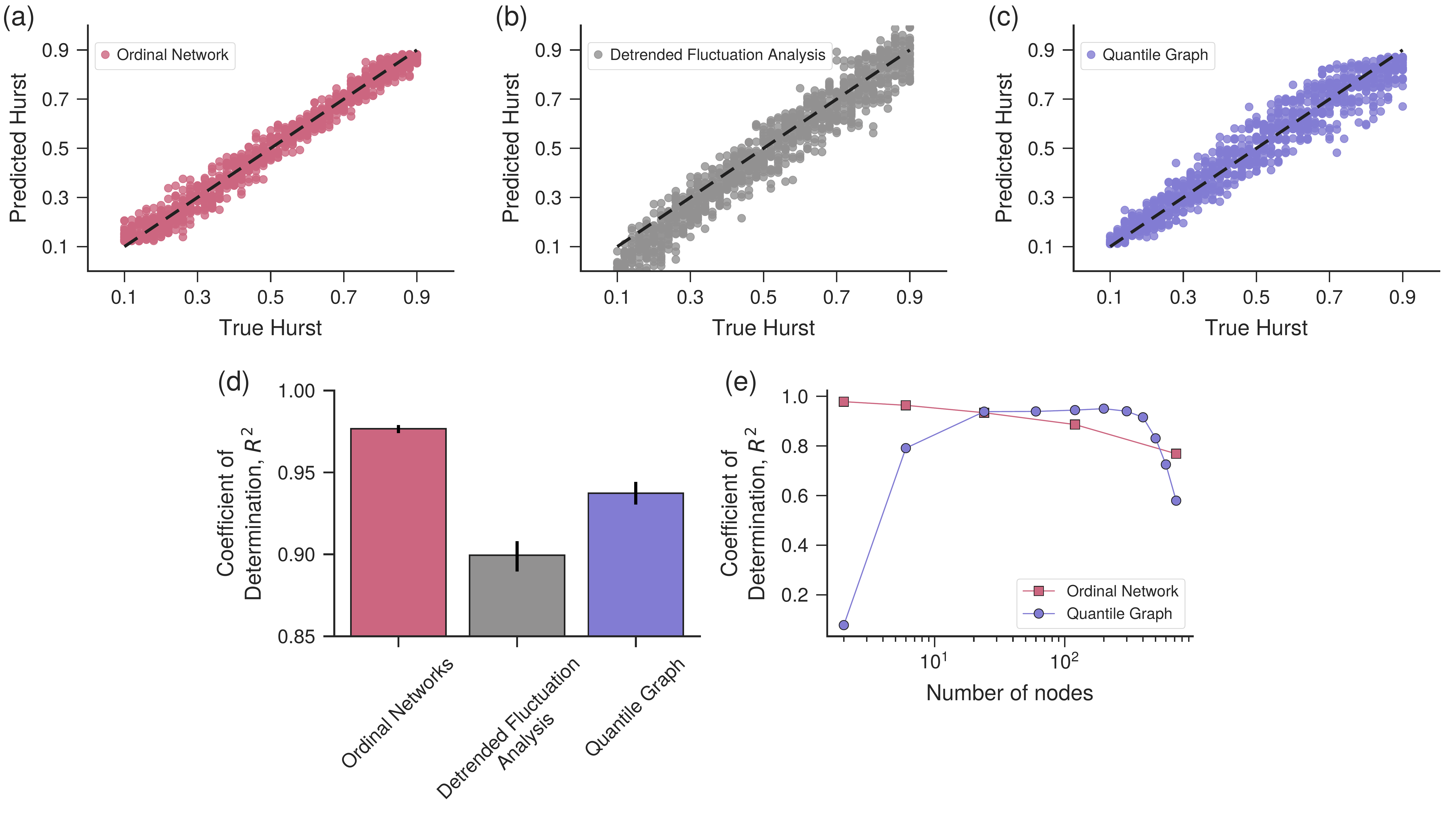}
\caption{Estimating the Hurst exponent with ordinal networks. Panels (a)-(c) show the relationship between true and predicted values of Hurst exponent obtained via ordinal networks ($d=2$), detrended fluctuations analysis (DFA with linear trend), and quantile graphs ($q=50$), respectively. The dashed lines represent 1:1 relationships. The predictions are obtained by applying the trained $K$-nearest neighbors regressor algorithm to the values of average weighted shortest paths $\langle l\rangle$ (for ordinal networks) and vertex degree (for quantile graphs) related to fractional Brownian motion series of $1024$ data points. For DFA, the value of $h$ is obtained by least squares fitting the relationship between the fluctuation function $F(s)$ and the scale parameter $s$ on a log-log scale (that is, $\log F(s) \propto h \log s$). (d) The bar plot shows the accuracy of each approach as measured by the coefficient of determination $R^2$. Error bars are $95\%$ bootstrapping confidence intervals with $1000$ resamples. (e) Dependence of $R^2$ on the number of nodes in ordinal networks (squares) and quantile graphs (circles).}
\label{fig:7}
\end{figure*}

We have also evaluated the average values of the normalized global node entropy $H_{\text{GN}}$ in comparison with the normalized permutation entropy $H$. \autoref{fig:6}(d) shows the results for fractional Gaussian noises. We note that the behavior of $H$ versus $h$ is more concave than $H_{\text{GN}}$ versus $h$, and the values of $H_{\text{GN}}$ display a much broader range of variation than those of $H$. This behavior is similar to what we have reported for noisy periodic signals [\autoref{fig:5}(b)] and provides further support to the hypothesis that the global node entropy has larger discriminating power than the usual permutation entropy. We have further calculated the average weighted shortest path $\langle l\rangle$ for the fractional Brownian motion as a function of the Hurst exponent $h$. \autoref{fig:6}(e) shows that the values of $\langle l\rangle$ monotonically decrease with $h$ for embedding dimensions $d=2$ and $d=3$ (fractional Gaussian noise displays a quite similar behavior).

The monotonic and well-defined behavior of $\langle l\rangle$ versus $h$ suggests that we can use the average weighted shortest path to predict the Hurst exponent. To systematically test for this possibility, we have built a statistical learning regression task, where the Hurst exponents of fractional Brownian motions are predicted using the values of $\langle l \rangle$ for $d=2$ as the unique covariate. We have generated another ensemble of $100$ samples of fractional Brownian motions of length $1024$ for each $h \in \{0.1, 0.12, 0.14,\dots,0.9\}$ via Hosking's method. By using this data set, we have trained a $K$-nearest neighbors regressor algorithm~\cite{introduction_statistical_learning} (see Appendix~\ref{Appendix:A}) with $75\%$ of all series and used a five-fold cross-validation approach to select the optimal number of neighbors $k$. The remaining $25\%$ of the series (that were never exposed to the learning algorithm) are used for testing the accuracy of the predictions. \autoref{fig:7}(a) shows the relationship between true and predicted values for the Hurst exponent, where an accuracy (measured by the coefficient of determination $R^2$) of $R^2=97.7\%$ is achieved. This represents a remarkable precision, particularly when considering that the series contain only $1024$ elements. To provide a baseline accuracy, we have compared the results obtained from ordinal networks with those from detrended fluctuation analysis (DFA)~\cite{peng_dnadfa_1994}, a widely used approach for estimating the Hurst exponent that is considered a cutting edge and reliable method~\cite{shao2012comparing}. \autoref{fig:7}(b) shows the relationship between true and predicted/estimated values of $h$ obtained by applying the DFA to $25\%$ of the same ensemble of series. We immediately note a larger dispersion of this relationship that can be quantified by the value of $R^2=89.9\%$. 

We have further compared the accuracy of the ordinal network approach with another time-series-to-network map known as quantile graphs ~\cite{campanharo_duality_2011,campanharo2016hurst}. Quantile graphs are weighted and directed networks where nodes represent a given number $q$ of quantiles of the empirical probability distribution associated with the time series, and links are drawn based on the temporal succession of quantiles related to each time series element. We have build quantile networks with the same data set used for ordinal networks and tested several network metrics (average vertex degree, average weighted shortest path, diameter, clustering coefficient) as predictive features of the Hurst exponent. We find that average vertex degree (in, out, or both combined) displays the best performance for this regression task. The average weighted shortest paths are problematic for quantile graphs due to the emergence of infinite distances that are either associated with more than one network component or with unreachable nodes through the directed paths. \autoref{fig:7}(c) shows the relationship between true and predicted values of Hurst exponents for quantile graphs with 50 quantiles, where the coefficient of determination is $R^2=93.5\%$. This accuracy is higher than the DFA but lower than the precision obtained with ordinal networks, as summarized in \autoref{fig:7}(d). 

\begin{figure*}[!ht]
\centering
\includegraphics[width=1\linewidth]{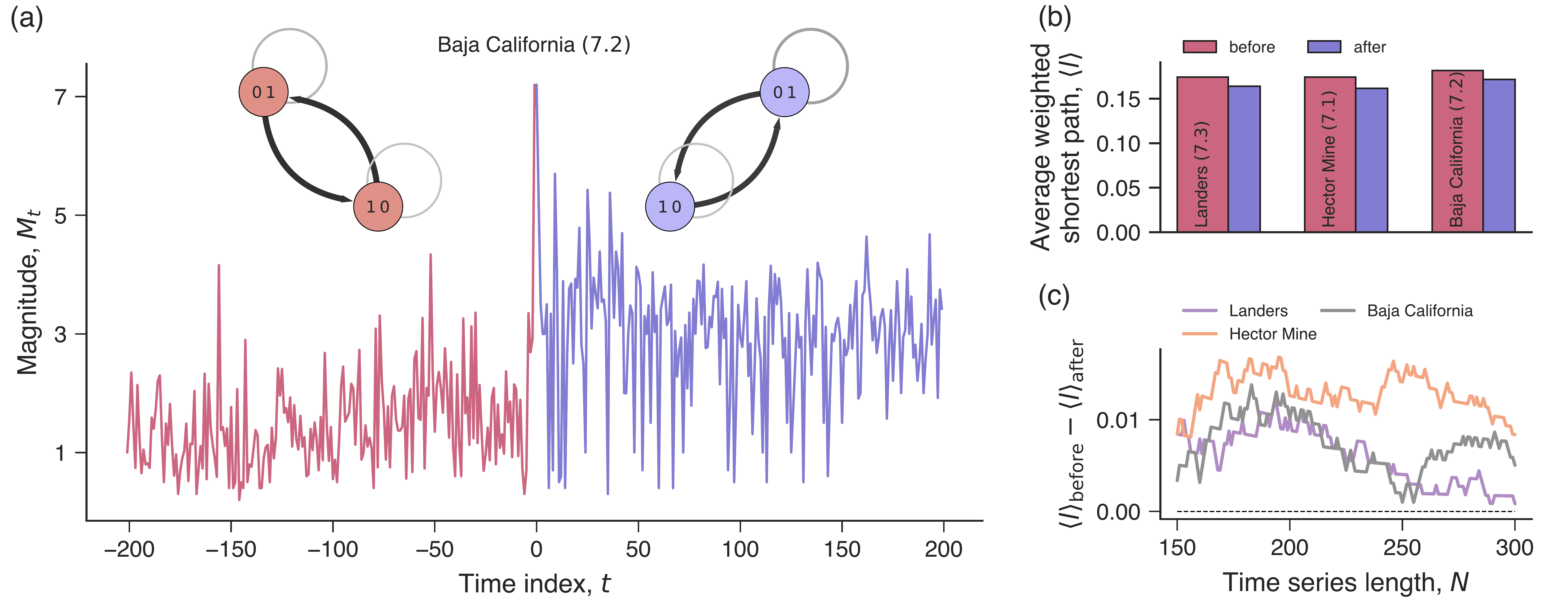}
\caption{Detecting changes in Earth seismic activity with ordinal networks. (a) Earthquake magnitude time series before (red) and after (purple) the great M $7.2$ Baja California earthquake occurred on April 4, 2010. The insets show a visualization of the corresponding ordinal networks with $d=2$ for the before and after time series. (b) Average weighted shortest path $\langle l \rangle$ estimated from ordinal networks built with before and after time series (200 events each) associated with three major earthquakes (indicated within the plot). We note that the values $\langle l \rangle$ decrease after the mainshock occurrence in the three cases. (c) Each colored curve shows the difference between the average weighted shortest path before ($\langle l \rangle_{\text{before}}$) and after ($\langle l \rangle_{\text{after}}$) as a function of the number of before and after events (that is, the time series length $N$). We observe that $\langle l \rangle_{\text{before}} - \langle l \rangle_{\text{after}}$ is always positive for $N$ within the range 150-300, which corroborates to the robustness of our results.}
\label{fig:8}
\end{figure*}

The number of quantiles in a quantile graph plays a similar role to the embedding dimension $d$ since both parameters define the number of nodes in the mapped network. Thus, we need to fine tune these parameters to have a fairer comparison between both approaches. To do so, we have trained the learning algorithms using different values of $d$ and the number of quantiles $q$, and estimated the values of $R^2$ in the test set for each combination. \autoref{fig:7}(e) shows $R^2$ as a function of the number of network nodes for both approaches. We observe that the increase of $d$ systematically reduces the performance of the regression task and that the optimum value occurs for $d=2$ [the same reported in \autoref{fig:7}(a)] for these time series with 1024 elements. On the other hand, quantile graphs show very low accuracy when the number of quantiles is too small ($q\lessapprox10$) or too large ($q\gtrapprox500$), exhibiting an optimal number of quantiles around $q\approx50$ [the same reported in \autoref{fig:7}(c)]. Therefore, we conclude that the performance of the regression task under optimized values of $d$ and $q$ is significantly better for ordinal networks ($R^2=97.7\%$ for $d=2$) than for quantile graphs ($R^2=93.5\%$ for $q=50$).

\subsection{Ordinal networks of earthquake magnitude series}\label{sec:results_earthquake}
As the last application, we have investigated ordinal networks emerging from time series of Earth seismic activity. In particular, we have analyzed earthquake magnitude time series from the Southern California Seismic Network~\cite{SCDC2019} between the years 1990 and 2019. We have asked whether ordinal networks are capable of detecting changes in the behavior of magnitude series caused by the occurrence of a large earthquake event (mainshock). To test for this possibility, we have selected all events of magnitude higher than 7.0 and built two time series composed of $N$ events before and $N$ events after the mainshock. \autoref{fig:8}(a) shows an example of before and after time series for $N=200$ related to the M\ 7.2 2010 Baja California earthquake that took place on April 4, 2010, at Guadalupe Victoria, a small city in the Mexican state of Baja California~\cite{Baja}. There were also two other large earthquakes in our data: the M\ 7.3 1992 Landers (June 28, 1992)~\cite{Landers} and M\ 7.1 1999 Hector Mine (October 16, 1999)~\cite{Hector}.

We have thus mapped the before and after magnitude series of those three mainshocks into ordinal networks with $d=2$ and $N=200$, as illustrated in \autoref{fig:8}(a). We calculate the average weighted shortest path $\langle l \rangle$ for the before and after networks. The results of \autoref{fig:8}(b) show that the values of $\langle l \rangle$ always decrease after the occurrence of a large earthquake. We have also verified that this result is robust against changes in the number of before and after events within the range $150$-$300$, as shown in \autoref{fig:8}(c). The decrease in $\langle l \rangle$ after the occurrence of a large mainshock is likely to be related to Omori's law~\cite{utsu1995centenary}, one of the fundamental seismic laws establishing that the number of aftershocks per unit of time decays as a power-law function of the elapsed time since the mainshock. We have verified that the Omori decay increases the persistence in the time series after mainshock events, as quantified by the lag-$1$ autocorrelation coefficient that increases from $0.1$ to $0.2$ (average values over the three events for $N=200$) after the mainshocks. This result thus indicates that the decrease in $\langle l \rangle$ may be associated with the increase in the persistent behavior of time series after mainshock events. It is also worth mentioning that the small length of these time series ($N\leq300$) prevents us from obtaining reliable estimates of the permutation transitions (and so of the ordinal networks) for higher embedding dimensions.

\section{Conclusions}\label{sec:conclusions}
We have presented an investigation of ordinal networks mapped from time series of stochastic nature. In particular, we have analyzed ordinal networks emerging from random series, noisy periodic signals, fractional Brownian motions, and earthquake magnitude sequences. We have provided a detailed description of random ordinal networks, revealing some counter-intuitive properties such as the non-uniform distribution of edge weights and the existence of auto-loops only in nodes related to solely increasing or decreasing permutations. We have also proposed an approach that is capable of building the exact form of random ordinal networks, which in turn are used for detecting non-random behavior in time series and missing permutation transitions. Our results for noisy periodic signals have indicated that the global node entropy estimated from ordinal networks is more robust against the presence of noise than the standard permutation entropy. We have further demonstrated the usefulness of ordinal networks for estimating the Hurst exponent of time series and for detecting sudden changes in earthquake magnitude series after the occurrence of large mainshock events.

We thus believe our work contributes for a better understanding of the general properties of ordinal networks, shedding light upon results and applications related to times series of stochastic nature.

\begin{acknowledgments}
This research was supported by Coordena\c{c}\~ao de Aperfeicoamento de Pessoal de N\'ivel Superior (CAPES) and Conselho Nacional de Desenvolvimento Cient\'ifico e Tecnol\'ogico (CNPq --- Grants 407690/2018-2 and 303121/2018-1).

\end{acknowledgments}

\appendix
\section{Statistical learning algorithm}\label{Appendix:A}

The $K$-nearest neighbors (KNN) is a supervised statistical learning algorithm used in both classification or regression tasks~\cite{introduction_statistical_learning}. The term supervised indicates that the algorithm analyzes a fraction of the data set (the training set) to produce an inferred function that is then used for predicting the behavior of data instances. In particular, the KNN algorithm determines the value of a new observation by averaging the values of the $K$ closest data points (using the space spanned by the independent variables in the training set). Thus, the number of nearest neighbors $K$ is a parameter of this algorithm, which is usually determined by simultaneously minimizing the bias and variance errors of the predictions~\cite{introduction_statistical_learning}. Bias errors happen when the learning model is too simple to represent an adequate description of the data. Variance errors, on the other hand, emerge when a complex model adjusts very well to the training set but is not able to generalize to unseen instances.

As mentioned in the main text, we have used the average weighted shortest $\langle l \rangle$ of ordinal networks and the average vertex degree $\langle k \rangle$ of quantile graphs in order to estimate the Hurst exponent $h$ of samples of fractional Brownian motion with the KNN algorithm. To do so, we generate an ensemble of 100 pairs of values $(h,\langle l \rangle)$ and $(h,\langle k \rangle)$ for each $h \in \{0.1, 0.12, 0.14,\dots,0.9\}$. We then randomly select $75\%$ of these data for training (training set) the KNN algorithm and let the remaining $25\%$ for testing the accuracy of the predictions. By using the training set, we have employed a five-fold cross-validation approach~\cite{introduction_statistical_learning} in order to determine the optimal value for the parameter $K$. This process consists of splitting the training set into $5$ subsets, using one of the subsets for validating the algorithm, and the remaining $4$ for training. This process is repeated $5$ times so that each subset is used for validation. In each step, we estimate the accuracy (here the coefficient of determination $R^2$) from the training subsets (the training score) and from the validating subset (the validation score). After $5$ repetitions, we estimate the average values of the training and validation scores and their confidence intervals. The plot of these scores as a function of the number of nearest neighbors $K$ (the so-called validation curves) allows us to identify the optimal value for this parameter, that is, the one corresponding to the highest value for the validation score. With this procedure, we find that the optimal value of $K$ for the results reported in the main text is $K=125$ for ordinal networks [\autoref{fig:7}(a)] and $K=77$ for quantile graphs [\autoref{fig:7}(c)].

\bibliography{ordinal_network}

\end{document}